\documentclass[12pt]{article}

\usepackage{graphicx}
\usepackage{amsmath,amssymb} 
\usepackage{tikz,multicol}
\usepackage{algorithm2e}
\usepackage{hyperref}

\def\cocoa{{\hbox{\rm C\kern-.13em o\kern-.07em C\kern-.13em o\kern-.15em A}}}

\newcounter{ejemplo}

\newtheorem{theorem}{Theorem}

\newtheorem{example}[ejemplo]{Example}

\newtheorem{lemma}[theorem]{Lemma}

\newtheorem{remark}[theorem]{Remark}

\newenvironment{proof}[1][Proof]{\textbf{#1.} }{\ \rule{0.5em}{0.5em}}

\providecommand{\keywords}[1]
{
  \small	
  \textbf{\textit{Keywords---}} #1
}

\title{Lasso for hierarchical polynomial models}
\author{H. Maruri-Aguilar\footnote{School of Mathematical Sciences, Queen Mary, University of London, London E1 4NS}
\footnote{\href{mailto:H.Maruri-Aguilar@qmul.ac.uk}{H.Maruri-Aguilar@qmul.ac.uk}}{ } and 
S. Lunag\'omez\footnote{Department of Mathematics and Statistics, Lancaster University}}

\begin{document}
\maketitle

\begin{abstract}
In a polynomial 
regression model, the divisibility conditions implicit in 
 polynomial hierarchy give way to a natural construction
of constraints for the model parameters. 
We use this principle to derive versions of strong 
and weak hierarchy and to extend existing work in the literature, 
which at the moment is only concerned with models of degree two.
We discuss how to estimate parameters in lasso using 
standard quadratic programming techniques and apply our
proposal to both simulated data and examples from the 
literature. The proposed methodology compares favorably
with existing techniques in terms of low validation error
and model size.
\end{abstract}


\section{Introduction}

This paper is concerned with polynomial regression models and
within this class of models, hierarchical polynomial models.
Our primary goal is to develop parameter constraints that 
enforce hierarchy for such linear models. 
In this paper we develop constraints for both strong and weak hierarchy,
using to our advantage the divisibility conditions of model
terms. 

A polynomial model is hierarchical when the presence of 
an interaction term such as $x_1x_2$ implies that both terms $x_1$ and 
$x_2$ are also in the model, in the sense that the coefficients of the
terms involved are not zero. This type of hierarchy is known in the 
literature as strong hierarchy \cite{BTT2013} and 
implies that hierarchical polynomial models must contain an 
intercept term. Weak hierarchy is a less restrictive form of 
hierarchy that has also been studied in the literature. In weak hierarchy,
the interaction term $x_1x_2$ would only vanish from the model when both
terms $x_1$ and $x_2$ have vanished. In this paper
the terms ``hierarchy'' and ``strong hierarchy'' are interchangeable, 
while ``weak hierarchy'' refers only to this type of hierarchy.

There are several arguments for the relevance of hierarchy in modelling. 
For instance, often the analysis 
is performed in linearly translated coordinates. If the model under
consideration in the transformed scale is not hierarchical, when translating 
back to the original scale, model terms that were not present in the transformed 
scale appear \cite{P1987}. 
Another case for hierarchy is that, e.g. were the intercept term be removed 
from the model, this would force it to pass through the origin. This type of 
model constraint should not be allowed to happen unless there is a strong 
reason for it \cite{MN1983}. 

Practical sparsity, the number of variables measured, is another argument 
used for hierarchy. 
Small models in the sense of low value of practical sparsity are achieved
through hierarchy. Indeed hierarchy ``reuses'' variables through 
interactions and higher order terms
 and should be preferred to modeling without hierarchy 
considerations \cite{BTT2013}.

There are several challenges for using and implementing hierarchy in models. 
A paramount challenge is to have a simple, consistent and 
intuitive way to formulate hierarchical models. As part of modeling and data 
analysis, a challenge is to estimate model parameters. Estimation methodologies 
 should be fast and 
efficient and the computational burden implied must be kept to a minimum. 

Another challenge is model 
size. Despite having relatively low practical sparsity, a hierarchical model may
still have many parameters.
 For example, if $k$ is the number of explanatory variables, 
a model with linear terms and squarefree interactions of order two
has $k(k+1)/2$ parameters excluding intercept, 
and $k(k+3)/2$ if pure quadratic terms are added. Once triple interactions are 
considered, the size of a full square free model 
is $(k+1)(k+2)(k+3)/6$. In short, model size can increase considerably depending on the number
of variables and the degree of terms used. Hence when using hierarchy there is a need to 
balance between the benefits of relatively big models and keeping the models to a manageable 
size.


\subsection{Hierarchy in the literature}

Our work is developed for a standard linear 
polynomial regression model 
\[Y = X\theta + \epsilon,\]
were $X$ is the design-model matrix, assumed to have full rank. It
has size $n\times p$ with $n>p$, and the $p$ columns of $X$ correspond to 
polynomial terms in $k$ explanatory variables, and $n$ is the 
number of observations.
The response vector is $Y$ 
and $\theta$ is the 
vector of model parameters, while $\epsilon$ is a vector
of independent error terms with zero mean and variance $\sigma^2$.

Polynomial models that satisfy strong hierarchy are also
known in the literature as ``well-formulated
polynomial regression models'' \cite{P1987,P1990}. 
Hierarchical models are also known in statistical 
literature as models that have the the property of
marginality, see \cite{MN1983,N1997,N1998,N2000}. Marginality,
that is hierarchy,
is routinely used for modeling in experimental
designs see \cite{B2008}.

There are several versions of polynomial hierarchy available in the statistical
literature and that have been implemented in \texttt{R} packages.
The authors in \cite{BTT2013} considered a second order 
polynomial model.
For this model, constraints were developed to achieve versions of strong and weak 
hierarchy. This proposal was developed to create the 
package \texttt{hierNet} which is Lasso for hierarchical
second order models. These ideas were later explored and 
developed further for hypothesis testing, see \cite{BST2015}.
Another development, termed \texttt{VANISH}, also considers a  
second order model as well as a functional extension of it, and they constructed 
a penalty that imposes strong hierarchy while keeping the criterion convex \cite{RG2010}.
In the development known as \texttt{FAMILY}, convex penalty functions are created using the 
rows and columns of the matrix of quadratic and second order interaction coefficients. This development
has also been implemented in \texttt{R} package  and considers different penalties 
that allow both cases of strong and weak hierarchy \cite{HWS2014}.

A two step hierarchical approach for the quadratic model is available in the package \texttt{glinternet} 
which first screens candidate main effects and interactions and implements group lasso to select 
variables while enforcing strong hierarchy \cite{LH2015}.
A recent contribution is the sequential search for hierarchical models while
simultaneously keeping low a notion of false rejections. This search 
has been implemented in the package \texttt{rai} and is potentially able to explore models
with higher 
order polynomial interactions \cite{JSF2019}.

A different approach for hierarchical polynomial model selection is that by \cite{BGW2003},
who sequentially search and discard model terms. A model is then selected with a compound 
criterion based on model curvature and validation error. This approach is not limited to 
polynomial models of second degree, however the search can be prohibitively expensive.

Finally, the authors in \cite{CZ2010} explored a general model parametrization that guarantees
hierarchy, but in the face of nonlinearity of this approach, they developed hierarchy in
a Bayesian context. Another proposal within the context of Bayesian analysis 
is \cite{NOY2015}. 

\subsection{Contributions}



Our first contribution is the development of general, non Bayesian 
methodology  for the analysis of data with hierarchical models. 
Our 
methodology is in practice as close as possible to standard lasso, while still 
enforcing hierarchy.
 Consider the plots in Figure \ref{fig_ex1d}, where the plot (a) is 
standard unconstrained lasso. The results for
constrained (b) are already hierarchical but still quite restrictive and we 
are able to ease the constraints to (c) and then (d) which is closer 
to (a) but keeping hierarchy.

Our contributed methodology has also low 
validation error and compares favorably with lasso and other existing methods from
the statistical 
literature. For example, in Figure \ref{fig_olive}, validation errors obtained for our 
hierarchical models (blue boxplots) improve over lasso and perform  
well when compared against other methods from literature.

Another contribution is the development of theory for our proposal. The methodology
we develop is based on Hasse diagrams. For a given candidate polynomial model, the 
divisibility conditions between terms can be encoded as a Hasse diagram. Therefore, 
such diagram serves to represent the constraints for  the parameter vector $\theta$ 
that guarantee hierarchy. The
use of Hasse diagrams was motivated by the experimental design literature, where
they are routinely used to analyze hierarchical models (see \cite{B2008}) but the
constraints are novel and to the best of our knowledge, such diagrams have not
been used previously to construct constraints.

In terms of parameter estimation,  we 
use lasso \cite{T1996}, which we adapt for hierarchy.
To guarantee hierarchy, the minimization of lasso 
criterion $L$  is constrained to the conditions on $\theta$ read from the Hasse diagram
and in practice we use an implementation of the quadratic programming methodology
by \cite{GI1983}.

\subsection{Order of the paper}

The order of the rest of the paper is as follows. 
In Section \ref{sec_hier} we define and then use the divisibility conditions 
implicit in a hierarchical polynomial model. We build a Hasse diagram from which
we read parametric constraints that guarantee model hierarchy. We then
discuss the generation of strong and weak hierarchies as well 
as the relation between such parametric constraints.
In Section \ref{sec_estimation}
we apply the constraints from Section \ref{sec_hier} as part of estimation in
 lasso. We develop this constrained estimation within standard lasso and also
develop a relaxation of it. 
In Section \ref{sec_ex} we apply our methods to examples from the literature.
We add a discussion in Section \ref{sec_disc} in which we comment on potential extensions
to the methodology.

\section{Hierarchical polynomial modelling}\label{sec_hier}

We first define polynomial notation, and elaborate on hierarchical 
polynomial models which satisfy divisibility conditions. Then using
Hasse diagrams, we 
develop the constraints on parameters
implied by hierarchy. The basic reference for polynomial notation is
\cite{CLO2007}, and for the use of this notation in statistics, see \cite{PRW2001}.

Consider $k$ indeterminates $x_1,\ldots,x_k$.
A monomial term $x^\alpha$  
is defined as the power product $x^\alpha:=\prod_{i=1}^kx_i^{\alpha_i}$, 
where  $\alpha$ is the exponent vector
$\alpha=(\alpha_1,\ldots,\alpha_k)$
whose entries are 
non negative integers. The degree of the term
$x^\alpha$ is the sum of its exponents
$\sum_{i=1}^k\alpha_i$.
Let $M$ be a finite set of exponent vectors so that 
the expectation of a linear regression
with terms in $M$ is
\begin{equation}\label{ec_model}\mbox{E}(Y(x))=\sum_{\alpha\in M}\theta_\alpha x^\alpha,\end{equation}
where $\theta_\alpha$ is the coefficient associated with
the term $x^\alpha$. Each coefficient $\theta_\alpha$ is a
fixed real quantity and thus the right hand 
side of Equation (\ref{ec_model}) is a polynomial. 
We refer to $M$ as the model, as it is the set of
candidate terms which will be used to model the response.


A model $M$ satisfies strong hierarchy 
when for every term $x^\alpha,\alpha\in M$, all the divisors of
$x^\alpha$ have exponents in $M$ as well. This is the strong hierarchy 
described in 
the introduction of this paper, and we  imply that the corresponding
coefficients are non zero.
The list of elements of a hierarchical $M$ can
be retrieved from the list of directing monomials of $M$, where
a directing monomial is a monomial $x^\alpha,\alpha\in M$ 
that cannot be divided by other monomial terms from $M$, see \cite{BGW2003}.

A polynomial model $M$ satisfies weak hierarchy,
when, for every term $x^\alpha$ with $\alpha\in M$, at least one of the divisors of
$x^\alpha$ have exponents in $M$ as well. In a model, strong hierarchy 
implies weak hierarchy.

\subsection{Hierarchy and partial ordering}
There is
a natural ordering of monomials implied by monomial
division and
denoted with the symbol $\prec$.
Consider two distinct monomial 
terms $x^\alpha,x^\beta$. We say $x^\alpha\prec x^\beta$ 
when $x^\alpha$ divides $x^\beta$ or
conversely, when $x^\beta$ is a monomial multiple of $x^\alpha$.
Note that $x^\alpha\prec x^\beta$ is attained when $\beta\geq\alpha$ 
componentwise, i.e. when $\beta-\alpha\geq 0$.

\begin{example}\label{exdiv}
Consider the set of monomials $\{1,x_1,x_2,x_3,x_1x_2,x_1x_3\}$ in
$k=3$ variables.
The divisibility condition $x_1\prec x_1x_2$ is equivalent to
checking componentwise that $\beta-\alpha=(1,1,0)-(1,0,0)=(0,1,0)\geq 0$.
The following are all the divisibility relations between monomials in the
list above:
$1\prec x_1$, $x_1 \prec x_1x_2$, 
$x_1 \prec x_1x_3$, 
$1\prec x_2 $, $x_2 \prec x_1x_2$, 
$1\prec  x_3$ and $x_3 \prec x_1x_3$. 
\end{example}

For a set of monomials with exponents from $M$, 
the ordering $\prec$ generated by divisibility is a transitive relation. An instance
of this in Example \ref{exdiv} is that 
$1\prec x_1$, 
$x_1 \prec x_1x_2$ so that $1\prec x_1x_2$ holds.
The ordering $\prec$
has guaranteed a unique minimum in $M$ only if the set $M$ includes $(0,\ldots,0)$,
i.e. the polynomial contains the intercept term $1$.
This is always the case when the model $M$ is hierarchical.
However note that $\prec$ defines only a partial order and not a total 
order in $M$. This is because in general, divisibility cannot 
 uniquely sort a list of monomials. A simple instance of this,
also taken from Example \ref{exdiv},
 is that $\prec$ per se cannot order $x_1$ from $x_2$. 
Despite the divisibility relation $\prec$ not being a total order, 
in this paper we do not require divisibility to satisfy this property and
to define hierarchy 
constraints in the model, it is sufficient to have 
a partial order.

\subsection{Hasse diagrams and model constraints}\label{sec_Hasse}

The collection of partial orderings among monomials which appear
by divisibility conditions
translates naturally into domination constraints for model
parameters. 
We next list those
relations $\prec$ in the model $M$ and later use the 
list to establish natural linear constraints 
between the model parameters.
Let $R$ be the collection 
of ordering relations between monomials in $M$:\begin{equation}R:=\{x^\alpha\prec x^\beta:
\alpha,\beta\in M \mbox{ such that } \beta-\alpha\geq 0 \mbox{ componentwise}\}.\end{equation}
We restrict $R$ to only 
contain those relations $x^\alpha\prec x^\beta$ when the degrees of
monomials $x^\alpha$ and $x^\beta$ differ only by one. 
The aim of this restriction is to keep $R$ to a 
minimum size by only listing essential relations, and
as $\prec$ is transitive, there is no lack of generality
by doing this.
The construction of the ordering relations $R$ for a model $M$ is given in 
Algorithm \ref{algoR}. 

All relations of the type $1\prec x_i$ are excluded from $R$
when the exponent $(0,\ldots,0)$ is removed from $M$ in the algorithm. 
Removing the intercept can be done because the intercept term is often of
little practical interest. Indeed after 
standardisation of data, the intercept term is absent from 
the modelling process. Unless stated otherwise, in the 
rest of this paper any relations or
constraints involving the 
intercept have been removed from both $M$ and the analysis.

The relations listed in $R$ can be depicted with a Hasse diagram 
\cite{BP2011}.
To build this diagram, model
terms are nodes and edges are drawn when terms are related by 
divisibility  $\prec$.  
Ascendant terms are divisors of other model terms
and they are
located at the top of the Hasse diagram. Conversely,
descendant terms can be written as polynomial multiples of other model
terms and they are located in the lower part of the diagram.
We put ascendant terms such as $x_1$ at the top of the Hasse diagram 
to reflect the importance of such terms in the parameter constraints 
to come later. 
The hierarchy of terms in the diagram we propose coincides 
with the hierarchy in Hasse diagrams when used in the 
analysis of experiments \cite{B2008}.

\begin{center}
\begin{algorithm}[H]
\SetAlgoLined\SetKwInput{KwData}{Input}\SetKwInput{KwResult}{Output}
\KwData{List of exponents $M$ }
\KwResult{List of divisibility conditions $R$}
 Initialization $M:=M\setminus \{(0,\ldots,0)\}$ and $R:=\{\}$\; 
 \For{$\alpha,\beta\in M$ with $\alpha\neq\beta$}{
  \If{$\beta-\alpha\geq 0$}{
	  \If{$\sum_{i=1}^k(\beta_i-\alpha_i)=1$}{
	     $R:=R\cup\{x^\alpha\prec x^\beta \}$\;
		} 
	 }
 }
 \caption{Generation of divisibility conditions $R$}\label{algoR}
\end{algorithm}
\end{center}

\begin{figure}
\begin{center}
\begin{tikzpicture}
  \node (a) at (-2,2) {$x_1$};
  \node (b) at (0,2) {$x_2$};
  \node (c) at (2,2) {$x_3$};
  \node (d) at (-1,0) {$x_1x_2$};
  \node (e) at (1,0) {$x_1x_3$};
  \draw (b) -- (d) -- (a) -- (e) -- (c);
  \draw[preaction={draw=white, -,line width=6pt}] (a) -- (e);
\end{tikzpicture}
\end{center}
\caption{Hasse diagram for model $\{1,x_1,x_2,x_3,x_1x_2,x_1x_3\}$.\label{ex_hasse}}
\end{figure}
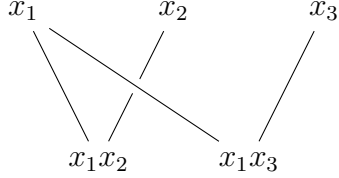

\begin{example}\label{exdiv2}
For the model $M$ of Example \ref{exdiv}, the set of
divisibility conditions is
$R=\{x_1\prec x_1x_2,
 x_1\prec x_1x_3,
x_2\prec x_1x_2,
 x_3\prec x_1x_3\}$.
These relations were used to build the Hasse diagram shown in Figure \ref{ex_hasse}.
\end{example}

A Hasse diagram can be built even when the model $M$ does not 
fully satisfy hierarchy. In such case the diagram would reflect the
hierarchical part of $M$ between terms that whose degree differ by one.

In summary, what we describe in this Section is model preprocessing, to be 
done before data analysis.
The model preprocessing does not involve response data, but only
applying Algorithm \ref{algoR} to model $M$ and then 
building Hasse diagrams and developing 
model constraints.

We now explore simple forms in which we can create and combine 
parameter constraints to obtain different forms of hierarchy.
These parameter constraints are read from the Hasse diagram built
for the model $M$ using the divisibility conditions encoded 
in the set $R$.

\subsubsection{Using the edges of the Hasse diagram}
The simplest way is to read parameter constraints directly from the 
edges in the Hasse diagram. For every relation $x^\alpha\prec x^\beta$ in
the list $R$, we associate the constraint 
$|\theta_\alpha|\geq |\theta_\beta|$ to the model. This constraint
ensures that if in the model the coefficient for $x^\alpha$ is zero,
i.e. $\theta_\alpha=0$,
then the term $x^\beta$ will be absent from the model as its coefficient
will be forced to satisfy 
$\theta_\beta=0$. 
Let $H$ be the 
list of all such constraints: 
\begin{equation}H:=\left\{|\theta_\alpha|\geq |\theta_\beta| 
\mbox{ for every pair }\alpha,\beta\in M\mbox{ such that }x^\alpha\prec x^\beta\in R \right\}.  
\end{equation}
The constraints imposed by $H$ correspond to strong hierarchy and there are as many 
constraints of this type as edges in the Hasse diagram.

If there is a model term $x^\alpha$ for which there are no divisors (ascendants)
nor multiples of it (descendants), then for such term there will be no 
divisibility conditions in $R$ and hence no parameter constraints appearing on 
the set $H$. The 
simplest example of this is when fitting a model with only linear terms so 
that, 
after discarding the intercept,
the expectation is $\mbox{E}(Y(x))=\sum_{i=1}^k\theta_i x_i$. The sets $R$ 
and $H$ are empty and estimation of model parameters involves no constraints.

\begin{example}\label{exdiv3} (Continuation of Example \ref{exdiv2})
From the Hasse diagram of Figure \ref{ex_hasse}, the following parameter 
constraints are directly read:  
$|\theta_1|\geq |\theta_{12}|$, $|\theta_1|\geq |\theta_{13}|$,
$|\theta_2|\geq |\theta_{12}|$ and $|\theta_3|\geq |\theta_{13}|$.
 These parameter constraints 
are associated to the model to ensure that it
remains hierarchical.
\end{example}

\subsubsection{Adding over descendant nodes in the Hasse diagram}
 
We can constrain by adding over multiples of model terms.
For every term $x^\alpha$ with exponent $\alpha\in M$, 
define $B(\alpha)$ as
the collection of exponents of multiples of $x^\alpha$, taken from the list $R$, that is 
$B(\alpha):=\{\beta\in M \mbox{ : }  x^\alpha\prec x^\beta\in R \}.$ 
To build the constraint, let $w_\alpha$ be a positive weight and 
add over coefficients of multiples of the monomial $x^\alpha$:
\begin{equation}w_\alpha|\theta_\alpha|\geq \sum_{{\beta}\in B(\alpha)} |\theta_{\beta}|.\label{strongh}\end{equation}
Recall that $B(\alpha)$ is the list of descendants of $x^\alpha$ in the
Hasse diagram, restricted to the immediate descendants. Hence if no 
monomials are below $x^\alpha$ in the diagram, then $B(\alpha)$ 
will be empty and no constraint is built for $\theta_\alpha$.
The constraint (\ref{strongh}) implies strong hierarchy, as the 
vanishing of $\theta_\alpha$ implies vanishing of all its descendant terms. Define 
$S$ to be the set of all such constraints, built over all terms that have
multiples (descendant terms)
\[S:=\left\{w_\alpha|\theta_\alpha|\geq \sum_{{\beta}\in B(\alpha)} |\theta_{\beta}|:\alpha\in M
\mbox{ such that }B(\alpha)\neq\emptyset \right\}.\]

The selection of the weight $w_\alpha$ can be arbitrarily made. There
are three obvious choices.
One is to let the weight of $|\theta_\alpha|$ in (\ref{strongh})
to be the number of descendants of $x^\alpha$, that is
$w_\alpha=|B(\alpha)|$, where $|B(\alpha)|$ is the 
cardinality of the set $B(\alpha)$.
This selection of $w_\alpha$
is equivalent to constrain by letting the absolute value of each parameter 
exceeding the mean of the absolute values of the parameters of its immediate descendants.
A second possibility is to let all the weights $w_\alpha$ in $S$ be equal to one,
i.e. $|\theta_\alpha|$ exceeds the sum of absolute values of descendants of $x^\alpha$.
The second instance is a more restrictive,
penalizing heavily parameters of higher order terms. 
A third possibility is to let the weight $w_\alpha$ exceed $|B(\alpha)|$. This
is a less restrictive form of strong hierarchy. The relation between these 
choices is discussed in Section \ref{sec_relation}.

\begin{remark}\label{remark_bien}
When the model $M$ corresponds to a polynomial of degree two, our formulation of strong 
hierarchy $S$ with weights $w_\alpha=1$ coincides with the strong
hierarchy as developed by \cite{BTT2013}.
\end{remark}

The symmetry constraint for strong hierarchy by
 \cite{BTT2013} 
is precisely equal to our  constraint $S$ with unit weights, which is a collection of 
inequalities of the form 
\[|\theta_j|\geq \sum_{i=1}^p|\theta_{ij}|=|\theta_{1j}|+|\theta_{2j}|+\ldots+|\theta_{jj}|+\ldots+|\theta_{pj}|,\]
where $j$ runs over all variables $j=1,\ldots,p$. If the model $M$ only includes double interactions but not quadratic terms in
the variables then the parameter $\theta_{jj}$ is absent from the above inequality. In both 
versions of it, the
correspondence between constraints of Remark \ref{remark_bien} holds.

\subsubsection{Adding over parent nodes in the Hasse diagram}
We can constrain parameters by adding over divisors of model terms. 
In a similar manner as above, for every term $x^\beta$ with exponent
$\beta\in M$, define $A(\beta)$ as the collection of exponents of ascendant
terms taken from the list $R$, formally
$A(\beta):=\{\alpha\in M \mbox{ and }  x^\alpha\prec x^\beta\in R \}.$
Let $w_\beta$ be a positive weight and constrain the parameter 
of $x^\beta$ by adding over its ascendants 
\begin{equation}\sum_{{\alpha}\in A(\beta)}|\theta_{\alpha}|\geq w_\beta|\theta_\beta|.\label{weakh}\end{equation}
The set $A(\beta)$ lists all terms that are higher up than $x^\beta$ in
the Hasse diagram, restricted to those immediate ascendants. Akin to
the earlier development, 
if no monomials are above $x^\beta$ in the Hasse diagram, then $A(\beta)$ 
is empty and no constraints would be created for $\theta_\beta$.
The constraint (\ref{weakh}) implies weak hierarchy, as the coefficient of term
$x^\beta$ would only vanish when all the coefficients of its ascendant terms
have vanished. Mirroring what was done earlier, define $W$ to be the set of all
constraints (\ref{weakh}), built over terms that have ascendants
\[W:=\left\{\sum_{{\alpha}\in A(\beta)}|\theta_{\alpha}|\geq |\theta_\beta|: 
\beta\in M\mbox{ such that } A(\beta)\neq\emptyset\right\}.\]

The specification of each weight $w_\beta$ is arbitrary and we also
consider three cases. The first is to let this weight to
be the number of ascendants, which we write
$w_\beta=|A(\beta)|$ and this means that the absolute value of the coefficient 
for $x^\beta$ is smaller than the mean of absolute values of its 
ascendants nodes.
The second case is to let all weights to be $w_\beta=1$. This
is less restrictive than the case above, being easier to attain. A third case
is to let weights $w_\beta$ be smaller than one, making the constraints much less
restrictive than the first two cases.
In the next example we give the explicit constraints for
the different cases of hierarchy.
\begin{example}\label{exdiv4} We build
different parameter constraints using the set $M$ 
of Examples \ref{exdiv} and \ref{exdiv2}.
The model
constraints appearing from the edges of the diagram are
\[H=\left\{|\theta_1|\geq |\theta_{12}|,\; 
|\theta_1|\geq |\theta_{13}|,\;
|\theta_2|\geq |\theta_{12}|\mbox{ and }
|\theta_3|\geq |\theta_{13}|\right\}.\]
Using weights $w_\alpha=1$, we add over 
multiples of every node in the diagram (adding over descendants) to obtain strong hierarchy $S$
\[\{ |\theta_{1}|\geq |\theta_{12}| + |\theta_{13}|,  \;\; |\theta_{2}|\geq |\theta_{12}| 
\mbox{ and }|\theta_{3}|\geq |\theta_{13}|\}\]
while with weights $w_\beta=1$ and adding over divisors of every node (ascendants) we have weak hierarchy $W$
\[\{ |\theta_{1}|+|\theta_{2}|\geq |\theta_{12}| \mbox{ and }|\theta_{1}|+|\theta_{3}|\geq |\theta_{13}|\}.\]
The version of strong hierarchy $S$ using weights $w_\alpha=|B(\alpha)|$ is
\[\{ 2|\theta_{1}|\geq |\theta_{12}| + |\theta_{13}|,  \;\; |\theta_{2}|\geq |\theta_{12}| 
\mbox{ and }|\theta_{3}|\geq |\theta_{13}|\},\] 
while weak hierarchy $W$ with weights $w_\beta=|A(\beta)|$ is
\[\{ |\theta_{1}|+|\theta_{2}|\geq 2|\theta_{12}| \mbox{ and }|\theta_{1}|+|\theta_{3}|\geq 2|\theta_{13}|\}.\]
\end{example}

\subsection{Relations between constraints}\label{sec_relation}

Consider the constraint 
$|\theta_{1}|\geq |\theta_{12}| + |\theta_{13}|$ taken from set of strong
hierarchy constraints $S$ of Example \ref{exdiv4}. It is clear 
that, if this constraint is satisfied,
it follows that both 
$|\theta_{1}|\geq |\theta_{12}|$ and 
$|\theta_{1}|\geq  |\theta_{13}|$ are also satisfied. This is
because 
$|\theta_{12}| + |\theta_{13}|\geq |\theta_{12}| $ and 
$|\theta_{12}| + |\theta_{13}|\geq |\theta_{13}| $ simultaneously. 
Thus, stemming from a constraint in $S$, we
have recovered some constraints from the set $H$. 
Lemma \ref{lemmaSH} gives the conditions under which 
the set of constraints $H$, built over the edges if the Hasse diagram,
can be deduced from the addition over descendants $S$.

\begin{lemma}\label{lemmaSH} If every weight $w_\alpha$ in the set of constraints $S$ satisfies 
$0<w_\alpha\leq 1$, then the constraints in $S$ imply the set of 
hierarchical constraints $H$.
\end{lemma}

Theorem \ref{th_implica}
establishes the relation between the sets of constraints for strong
hierarchy $S$ and weak hierarchy $W$. The implications of the
theorem depend on the weights specified in each case. The proof of
both results is in the Appendix.

\begin{figure}
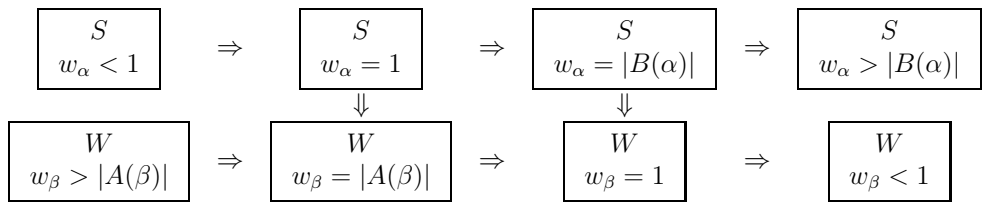

\begin{center}\scalebox{0.85}{
\begin{tabular}{ccccccc} 
\fbox{\begin{tabular}{c}$S$\\$w_\alpha<1$\end{tabular}}&
 $\Rightarrow$&
\fbox{\begin{tabular}{c}$S$\\$w_\alpha=1$\end{tabular}}&
 $\Rightarrow$&
\fbox{\begin{tabular}{c}$S$\\$w_\alpha=|B(\alpha)|$\end{tabular}}&
 $\Rightarrow$&
\fbox{\begin{tabular}{c}$S$\\$w_\alpha>|B(\alpha)|$\end{tabular}}\\
&&$\Downarrow$&&$\Downarrow$\\
\fbox{\begin{tabular}{c}$W$\\$w_\beta>|A(\beta)|$\end{tabular}}&
 $\Rightarrow$&
\fbox{\begin{tabular}{c}$W$\\$w_\beta=|A(\beta)|$\end{tabular}}&
 $\Rightarrow$&
\fbox{\begin{tabular}{c}$W$\\$w_\beta=1$\end{tabular}}&
 $\Rightarrow$&
\fbox{\begin{tabular}{c}$W$\\$w_\beta<1$\end{tabular}}
\end{tabular}}
\caption{Implications between constraints.}\label{fig_relations}
\end{center}

\end{figure}

\begin{theorem} \label{th_implica}Let the sets of constraints $S$ and $W$ be as
defined in Section \ref{sec_Hasse}. Then, depending on the
specification of weights $w_\alpha$ and $w_\beta$ for sets of constraints
$S$ and $W$, the implications shown in the diagram in Figure \ref{fig_relations} hold.
\end{theorem}

In the top row of the diagram in Figure \ref{fig_relations}, we have the constraints associated 
with strong hierarchy $S$, while the bottom row has constraints 
associated with weak hierarchy $W$. In the same diagram, the constraints are also ordered 
from left to right from the most restrictive instances to the least restrictive. The choice of
weights
provides the modeler with a wide range of models, all of which would satisfy the required
type of hierarchy.

Finally, we must distinguish between the hierarchy as built with parameter constraints that 
we have related in this section and hierarchy when built with arbitrary values of the parameters. 
While it is true that if a model with exponent set 
$M$ is strongly hierarchical then it is also weakly hierarchical, it is also true that 
the parameters associated with such models may not necessarily satisfy the implications 
of the theorems in this section. Indeed it is 
possible to have a model that is strongly hierarchical that does not satisfy any of 
the sets of constraints $S$ as defined in this paper. Our contribution
provides a useful set of parameter constraints that guarantee model hierarchy, and while it does not cover 
all possible cases, it gives a general and flexible modeling framework.

\section{Parameter estimation}\label{sec_estimation}

The development below is based around the lasso shrinkage
methodology \cite{T1996}.
This methodology estimates $\theta$ by minimizing over $\mathbb{R}^p$ the criterion
\begin{equation}\label{eclasso}
L=\frac{1}{2}||Y-X\theta||_2^2+\lambda||\theta||_1.
\end{equation}
If no shrinkage is needed, set $\lambda=0$ in $L$ to coincide
with least squares estimation.
We next discuss constrained lasso and then a relaxed version 
 of it. 

\subsection{Constrained lasso}\label{sec_classo}
The estimation problem for $\theta$ is to minimize $L$ of Equation (\ref{eclasso}) over $\mathbb{R}^p$
subject to a set of constraints, and we refer to this as constrained lasso.
In our proposal, the constraints are given by
one of the sets $S$, $H$ or $W$, selected by the modeler. With a slight abuse of notation, let $|\theta|$ denote the column vector 
whose entries are absolute values of coefficients, i.e.  $|\theta|=(|\theta_\alpha|)_{\alpha\in M}$. The 
parameter constraints take the form\[A|\theta|\geq 0,\] where 
$A$ is a matrix of constants with $p$ columns, read from the earlier Hasse development of Section \ref{sec_Hasse},
and the inequality is interpreted componentwise.

\begin{example} Consider the hierarchy $H$ as developed in Example \ref{exdiv4}, 
and let the vector of absolute values of coefficients be $|\theta|=
(|\theta_1|,|\theta_2|,|\theta_3|,
|\theta_{12}|,|\theta_{13}|)^T$, then the matrix $A$ for
the constrained optimization would be
\[A=\left(\begin{array}{rrrrr}1&0&0&-1&0\\ 1&0&0&0&-1\\
0&1&0&-1&0\\0&0&1&0&-1\\
\end{array}\right).\] 
For a given vector $|\theta|$, simultaneous attainment of all the inequalities
means that $|\theta|$ satisfies $H$.
For the cases of strong hierarchy S with weights $w_\alpha=|B(\alpha)|$ and for
unit weights, the matrix $A$ would be 
\[\left(\begin{array}{rrrrr}2&0&0&-1&-1\\0&1&0&-1&0\\0&0&1&0&-1\\
\end{array}\right) \mbox{ and } 
\left(\begin{array}{rrrrr}1&0&0&-1&-1\\0&1&0&-1&0\\0&0&1&0&-1\\
\end{array}\right),\] 
respectively. For weak hierarchy W with weights $w_\beta=|A(\beta)|$
and for unit weights, the matrices $A$ would be
\[\left(\begin{array}{rrrrr}1&1&0&-2&0\\1&0&1&0&-2
\end{array}\right) \mbox{ and }
\left(\begin{array}{rrrrr}1&1&0&-1&0\\1&0&1&0&-1
\end{array}\right).\] 
\end{example}

For a given value of $\lambda$,
the numerical minimization of constrained lasso is a  minimization of a quadratic form with absolute constraints. Note 
that for every orthant of $\mathbb R^p$, the minimization remains a standard quadratic problem with linear constraints. This 
is because inside every orthant, the vector of absolute values $|\theta|$ is linear with respect to each of its coordinates $\theta_\alpha$. 

In practice we use the following minimization procedure: select an initial orthant based on the least squares estimate, 
then for a set of values of $\lambda$, compute  lasso estimates in the selected orthant.
The minimization is done using the standard \texttt{R} package \texttt{quadprog} and by default, the values 
of $\lambda$  range from $\lambda=0$ to $\lambda=\max\{ |(X^TY)_i| \}$.
As part of our search procedure, we  
also explore if neighboring orthants given by a sign change may have a 
better estimate than our selected orthant and thus for every value of $\lambda$ we explore $|M|+1$ orthants. 
We do not claim that our search is a universally optimal procedure, but it has worked well in practice and it is much cheaper than a brute force exploration of
all $2^{|M|}$ full dimensional orthants of $\mathbb R^p$.

\subsection{Relaxed constrained lasso}\label{sec_relaxlasso}

An alternative approach to $L$ is to do
 a convex relaxation of the lasso problem, see \cite{F1981}.
This relaxation is used to linearize the sum of absolute values $||\theta||_1$ in
Equation (\ref{eclasso}). 
In the relaxed lasso, instead of the parameter 
vector $\theta$, we
have two non-negative parameter vectors $\theta^+$ and $\theta^-$, which for simplicity we collect 
in the column vector $u$ which has $2p$ rows and is $u^T=({\theta^+}^T,{\theta^-}^T)$. The
non-negative condition of $u$ from the relaxation implies that $u\in\mathbb{R}^{2p}_{\geq0}$.
The vector of parameters $\theta$ is built as the difference $\theta=\theta^+-\theta^-$, that is
\begin{equation}\theta=\left(\begin{array}{cc}I&-I\end{array}\right)u,\label{ecthetau}\end{equation}
where $I$ is an identity matrix of size $p$.  
The proxy vector of absolute values of $\theta$ is defined as
the addition of these nonnegative 
vectors \[|\theta|=\theta^++\theta^-=\left(\begin{array}{cc}I&I\end{array}\right)u\] so that
$||\theta||_1$ is replaced by the addition of all the elements
in $\theta^+-\theta^-$, that is by $u^T\mathbf{1}$ with $\mathbf{1}$ a column of ones
with $2p$ rows.
After collecting
terms, the relaxed version of the Lasso criterion is
\begin{equation}\label{eclassor}
L_r=\frac{1}{2}Y^TY-u^T\left( {X^TY\choose -X^TY} -\lambda \mathbf{1} \right)
+\frac{1}{2}u^T\left(\begin{array}{rr}X^TX&-X^TX\\-X^TX&X^TX\end{array}\right)u.
\end{equation}
The problem of estimation of $\theta$ minimizing (\ref{eclasso}) has been turned
into minimization of the relaxed criterion $L_r(u)$ of 
(\ref{eclassor}) subject to $Bu\geq 0$ (component wise). 
The matrix $B$ contains the constraints imposed by hierarchy as discussed earlier
as well as the non negativity constraints for values of $u$. 
The constraints matrix $B$ has the following block form \[B=\left(\begin{array}{cc}A&A\\
I&0\\0&I\end{array}\right),\] where $A$ is a matrix with $p$ columns that has the 
parameter constraints as developed in Section \ref{sec_Hasse}; the matrix $I$ is 
an identity matrix of 
size $p$ and in both cases above the matrix $0$ is of size $p\times p$.
Note
that when there are no edges in the Hasse diagram, then the matrix $A$ does not
exist and matrix $B$ for the relaxation of Lasso only contains the lower part with 
identity and zero matrices.

The estimation of the lasso path for the relaxed constrained lasso is performed in 
a similar manner to the procedure in Section \ref{sec_classo}: for a collection
of values of $\lambda$, estimates are computed using the same \texttt{R} package
\texttt{quadprog}.
Note that an advantage of the relaxation is that there is no need to explore
different quadrants, as the nonnegativity of $u$ and hierarchy constraints 
are all handled by $Bu\geq 0$.

We briefly discuss the construction of $B$ through an example.
Consider the model $M$ 
of Example \ref{exdiv} and 
let $\theta^T=(\theta_1,\theta_2,\theta_3,\theta_{12},\theta_{13})$ be
the parameter vector
so that for the relaxation we have
${\theta^+}^T=(\theta_1^+,\theta_2^+,\theta_3^+,\theta_{12}^+,\theta_{13}^+)$
and ${\theta^-}^T=(\theta_1^-,\theta_2^-,\theta_3^-,\theta_{12}^-,\theta_{13}^-)$
and $u^T=({\theta^+}^T,{\theta^-}^T)$.
Any constraints for this model that involve absolute values, such as those given 
in Example \ref{exdiv4}, are reformulated using the component wise
convention $|\theta_\alpha|=\theta_\alpha^++\theta_\alpha^-$. 
For instance, when considering hierarchy $H$
in the relaxed Lasso for model $M$,
the constraint $|\theta_1|\geq |\theta_{12}|$ is replaced  by
$\theta_1^++\theta_1^-\geq \theta_{12}^++\theta_{12}^-$. This
constraint is
rearranged  as 
$(\theta_1^+-\theta_{12}^+) +(\theta_1^-- \theta_{12}^-)\geq0$, where 
the brackets separate components of $\theta^+$ from those of $\theta^-$. 
Note the repetition of roles of elements of $\theta^+$ and of $\theta^-$ inside 
each bracket. Rearranging the inequality, we read the constraint
$(1,0,0,-1,0,1,0,0,-1,0)u\geq0$, where the duplication of roles in terms in brackets above 
implies the repetition of the coefficients as noted earlier.
The rest of the constraints in $H$ give the 
hierarchy constraints 
\[\left(\begin{array}{rrrrrrrrrr}1&0&0&-1&0&1&0&0&-1&0\\ 1&0&0&0&-1&1&0&0&0&-1\\
0&1&0&-1&0&0&1&0&-1&0\\0&0&1&0&-1&0&0&1&0&-1\\
\end{array}\right)u\geq 0,\] 
where, as in earlier developments, the inequality is interpreted component wise. This is simply a case
of 
\begin{equation}\left(\begin{array}{cc}A&A\end{array}\right)u\geq 0\end{equation}
where the matrix $A$ is the same as the development in the text of Section \ref{sec_classo}.
The nonnegativity constraints for $\theta^+$ and for $\theta^-$ are
simply written as 
\begin{equation}\left(\begin{array}{cc}I&0\end{array}\right)u\geq 0 \mbox{ and }
\left(\begin{array}{cc}0&I\end{array}\right)u\geq 0,\end{equation} 
respectively. In summary, for the
hierarchy $H$ in model $M$, the relaxed lasso uses a constraints matrix $B$ of size
$14\times 10$; the first four rows impose the hierarchy $H$, and the
remaining ten rows are an identity of size ten that gives non-negativity 
constraints for the relaxed parameterisation.

Relaxed constrained lasso appears to be a simple alternative to constrained lasso of Section \ref{sec_classo}. The relaxed 
method does not have to consider multiple orthants and move around them, as this is automatically handled by the problem formulation. 
Hierarchy is rigorously held for the vector of proxy absolute values $\theta^++\theta^-$, but paradoxically, the method 
does not guarantee that hierarchy will be enforced for the vector of proxy parameters $\theta^+-\theta^-$, nor that the model retrieved will
coincide with that of constrained lasso, see the numerical comparison of methods in Section \ref{sec_compara}.
 We would advocate this simple, approximate method for cases where strong hierarchy is not a crucial requirement, 
or for a first stage screening of model terms where only 
a rough list of active terms is required.

\section{Examples}\label{sec_ex}

\subsection{Small example, synthetic data} \label{ex_1d3}

\begin{figure}
    \centering
    \includegraphics[scale=0.75]{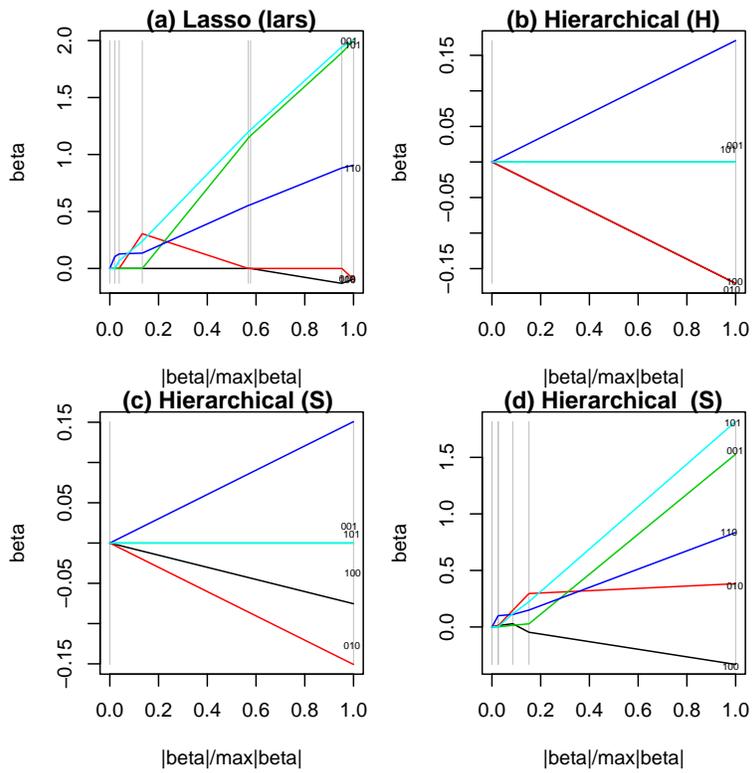}
    \caption{Lasso paths for Example \ref{ex_1d3}: (a) standard Lasso and hierarchical according to
    (b) $H$; (c) $S$ with $w_\alpha=|B(\alpha)|$ and
    (d) $S$ with $w_\alpha=8$. The colors of coefficients
    of terms $x_1,x_2,x_3,x_1x_2,x_1x_3$ in the path are black, red, green, dark blue and cyan, respectively.}
    \label{fig_ex1d}
\end{figure}

Consider the data set in $k=3$ factors given in Table \ref{tab_ex1d}. The model
under consideration has terms for the intercept and all linear
factors together with the interaction terms $x_1x_2$ and $x_1x_3$. 
Figure \ref{fig_ex1d} shows the lasso paths for standard lasso in (a) as well as the lasso paths  built using the following constraints as defined in this paper: $H$ in (b), $S$ with weights given by $w_\alpha=|B(\alpha)|$ in (c), $S$ with all weights equal $w_\alpha=8$ in (d). 

\begin{table}\begin{center}
\begin{tabular}{rrrr}
$X_1$&$X_2$&$X_3$&$Y$\\\hline
0&-1&-1&-2\\ -1&0&0&0\\ -1&-1&-1&1\\
-1&0&1&1\\ -3&-1&1&-1\\ -1&0&1&-1\\ 7&3&-1&2\\
\end{tabular}\end{center}
\caption{Simulated data for Example \ref{ex_1d3}.}\label{tab_ex1d}
\end{table}

The lasso path in Figure \ref{fig_ex1d} (a)
was computed with the standard package \texttt{lars} and added for reference, noting that the coefficients in this path do not obey hierarchy. The rest of the paths in Figure \ref{fig_ex1d} were computed minimizing $L$ subject to absolute constraints.
We note that the introduction of hierarchy $H$ severely constraints the lasso path to the point that the coefficient estimates for terms $x_3$ and $x_1x_3$ become zero for all the path and the
paths of $x_1$ and $x_2$ are fully correlated. This is of course too limiting, as seen in (b) in the said figure. The path of hierarchy $S$ shown in (c) is still limiting as
the same coefficients as in $H$ are still zero, but the gradual change in the path towards lasso starts to become evident. Finally, the path in (d) is much closer to the original lasso, while still keeping hierarchy.

\subsection{Prediction comparison: strong hierarchy and lasso}

We carried out a simulation study to compare the 
performance of our constrained lasso $S$ proposal versus
standard lasso. 
To this end, we considered data from a hierarchical
model assumed true with $11$ terms (excluding
intercept) in $k=2$ variables, and 
 integer coefficients taken from the interval $[-3,3]$
and an error term was added to the deterministic part
of the model. The model had  directing monomials $x_1x^3_2, x^2_1x^2_2$ and $x^3_1$.

The above simulated data was modelled with a candidate model with
$p=24$ terms and directing monomial $x^4_1x^4_2$.
The design used was a random uniform design in $[-1, 1]^2$  
with 
$100$ training, $40$ validation and $40$ prediction points.
After training, the validation error was used to pick a model 
from the path for each of the trajectories of lasso and constrained lasso $S$.
Finally, using the selected model and the prediction data,
we computed the prediction error and compared for both models.
This comparison was repeated $1000$ times for each of $18$ levels of
variance of Gaussian error, and for nine different
values of weights $w_\alpha$ ranging from $w_\alpha=1$ to $w_\alpha=100$. 

\begin{figure}
    \centering
    \includegraphics[scale=0.6]{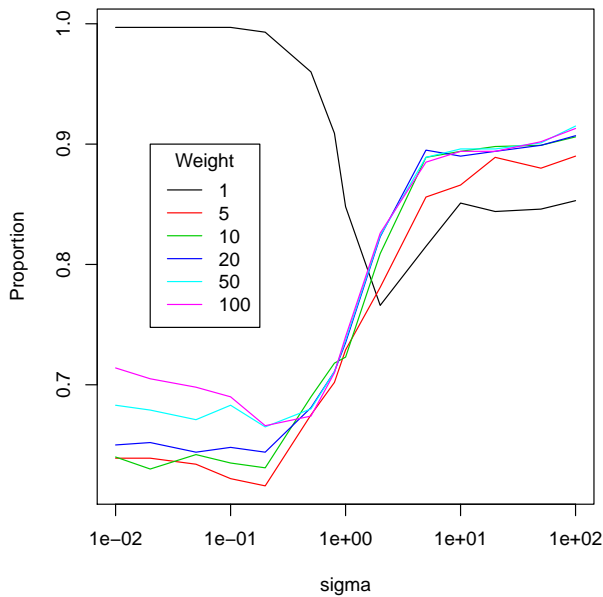}
    \caption{Proportion of times that error from strong hierarchy equaled or outperformed lasso
		error.}
    \label{fig_ex2d}
\end{figure}

Figure \ref {fig_ex2d} shows the proportion 
of times that the prediction error of constrained lasso $S$ was smaller
than or equal to that of lasso, plotted against weight $w_\alpha$. Firstly, in all simulations, this
proportion was at least $60\%$, and for increasing levels of error 
variance, the figure settled at around $85\%$. An interesting case 
appeared for lower values of the weight $w_\alpha$, for which the proportion
was initially 
much higher than the rest of cases, to finally settle for a similar
proportion as the rest of cases. 

\subsection{Benchmarking techniques with engine data set}

This data set was generated by a computer experiment involving $5$ input variables 
and one output. The first $48$ observations of the data were used
to train the model and the remaining $49$ observations of the
data were used for validation purposes. Two initial models were considered
for analysis.


In the original analysis by Bates et al. (2003), they considered a saturated model
of $48$ terms.
For our analysis, in order to have a non-saturated initial model,
we removed the single term of degree five in their model. This initial
model is referred to as `Initial BGW'.
Using the 
training data, we built 
constrained lasso paths and for each model in the path, we reestimated the
coefficients using least squares estimates. 
We then selected a model from the path using the validation error. This analysis was performed 
for hierarchy $H$ and several choices of weights for hierarchies $S$ and $W$. 

\begin{table}
\begin{center}
\begin{tabular}{l|ccc|ccc}\hline\hline
Initial model&\multicolumn{3}{c}{BGW}&\multicolumn{3}{|c}{Quadratic}\\\hline
Method (weight)&Size&MSE&Hier.&Size&MSE&Hier.\\\hline\hline
$S$ ($w_\alpha=1$)&21&1.3788&Y&13&1.6124&Y\\
$H$ &22&1.4575&Y&\;\;9&1.7124&Y\\
$S$ ($w_\alpha=|B(\alpha)|$)&21&1.5171&Y&12&1.6142&Y\\
$S$ ($w_\alpha=100$)&21&1.3476&Y&12&1.6142&Y\\
$W$ ($w_\beta=|A(\beta)|$)&20&1.6501&N&10&1.6756&N\\
$W$ ($w_\beta=1$)&22&1.4849&N&10&1.6756&N\\
Lasso&19&1.6179&N&10&1.6756&N\\
BGW&11&1.5742&Y&\textemdash&\textemdash&\textemdash\\
RAI&\textemdash&\textemdash&\textemdash&\;\;6&2.0909&Y\\\hline\hline
FAMILY&21&2.2487&Y&\textemdash&\textemdash&\textemdash\\\hline\hline
\end{tabular}
\end{center}
\caption{Validation error for the selected models, engine emissions data.}\label{table_rr}
\end{table}

The models, sizes and validation mean squared errors for this study are given in
Table \ref{table_rr}, and the last column of the table states whether the model
satisfies strong hierarchy. 
The constrained lasso models in the first six rows of Table \ref{table_rr} are 
ordered according to a reading from left to right by row in the diagram of 
Theorem \ref{th_implica}. The first four rows correspond to 
strongly hierarchical models which are ordered in decreasing order of
restriction. 
For example, the large weight $w_\alpha=100$ in row four 
made this model considerably less restrictive than any model above it, while still 
obeying strong hierarchy. The fifth and sixth row correspond to weak hierarchy
are also ordered in decreasing order of restriction. Table \ref{table_rr} 
also shows in its last three rows the results from
a standard lasso path, from the final model
by \cite{BGW2003}, termed BGW; as well as the model termed RAI from the
greedy stepwise regression search by \cite{JSF2019}.

A common feature in all the models we tried was that the term involving
the second variable was absent from the selected model, a desirable analysis consequence
noted in \cite{BGW2003}. From the models, the best result was obtained with
strong hierarchy $S$ and the large weight $w_\alpha=100$, which is 
the least restrictive of our constrained lasso with strong
hierarchy. The second best model is also a strongly hierarchical. 
Interestingly enough, the entirely unconstrained standard lasso came
second to last with its only benefit being a slightly smaller model size.
The model BGW, despite having a relatively large MSE in
Table \ref{table_rr}, has a reduced size. This is a consequence
of that model being obtained through a 
compound criterion that penalized heavily higher order terms by their 
curvature, in contrast with 
our methods that penalize higher order terms concerning hierarchy.
The smallest model was RAI, with the trade-off of having the largest 
validation error in the table.

\subsection{Comparison between methods of constrained estimation}\label{sec_compara}

We compared the performance of constrained lasso against the relaxed version
of it using simulated data of the non-polynomial model
$y = 1 + 2\exp(x_1) + 3\sin(\pi x_3)^2$, suggested by \cite{BGW2003}. We modeled and compared two
scenarios with $k=3$ and $k=5$ variables. In the first scenario, the data 
does not depend on one input factor $x_2$ while on the five dimensional 
scenario, factors
 $x_2,x_4$ and  $x_5$ have no influence in the output.

\begin{figure}
    \centering
    \includegraphics[scale=0.6]{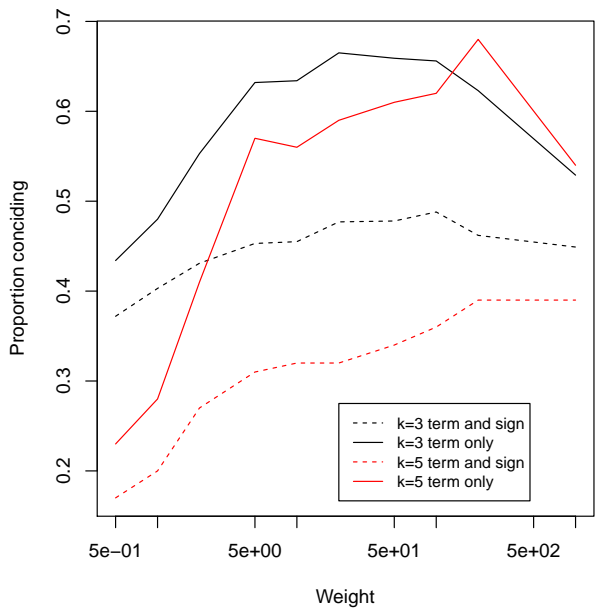}
    \caption{Proportion of times final relaxed models coincided with those of constrained
		lasso.}
    \label{fig_excompara}
\end{figure}

For the three factor scenario, we generated a random latin hypercube design (LH) of $40$ points
which were used to train a model of degree three with $19$ terms.
A second latin hypercube of $15$ points was
used to compute validation error which selected a working model. We did this for
constrained lasso and relaxed lasso with constraints, using the strong hierarchy $S$.
We recorded the proportion of times that the final model of relaxed lasso concided 
with that of the constrained version in both terms and respective coefficient signs. We also
recorded the proportion when they coincided on the terms only, disregarding the signs
of coefficients. These proportions were computed using $1000$ replications of this experiment, 
performed for a variety of weights $w_\alpha$ for constraint $S$.

For the five factor scenario, we used a LH design of $100$ points in five dimensions
to train
a model of $55$ terms with the same degree three and hierarchy $S$ as the previous 
scenario. The validation
design consisted of a second  LH of $30$ points, and we recorded the same proportions
as the earlier case, using instead $100$ replications.

Figure \ref{fig_excompara} shows the experiment results. For moderate to large values
of the weight $w_\alpha$, the coincidence of terms reaches a value of around $60\%$
for both three and five-dimensional scenarios. Coincidence of terms and signs is much
lower for both scenarios, never reaching even $50\%$ of the cases. The results suggest that 
while the relaxed lasso has the ability to detect active terms that obey the required 
hierarchy, it does not do that with a very high probability.

\subsection{Run times of the calculations}

\begin{figure}
    \centering
    \includegraphics[scale=0.6]{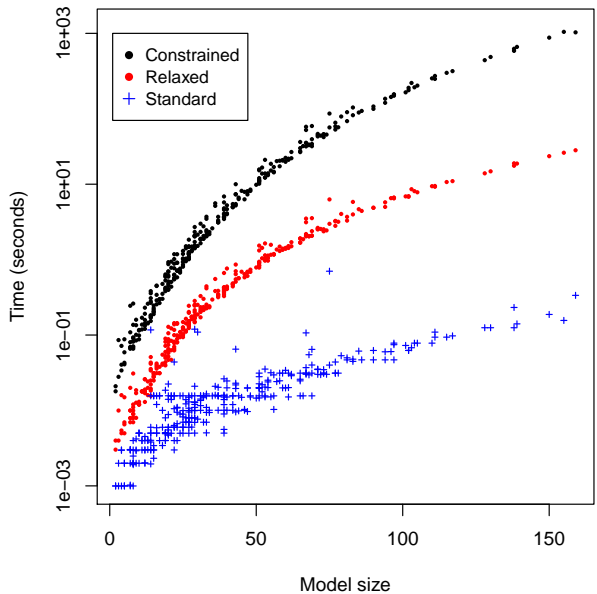}
    \caption{Run time for constrained $S$, relaxed $S$ and standard lasso.}
    \label{fig_runtimes}
\end{figure}

This example is concerned with a comparison of the run times of the implementation of 
our computational procedure. We randomly generated data scenarios varying numbers
of input factors and candidate models $M$ and to which
each of the constraints $S,H$ and $W$ was fitted to generate lasso paths in both versions
of constrained lasso and as 
relaxed lasso. Run times for a standard laptop (64 bit processor 1.8GHz, 8GB RAM) were recorded for each scenario and method.

The details of the simulation are as follows. The number of factors $k$ ranged 
from $1$ to $19$. For each value of $k$ we 
generated a random latin hypercube with $n$ points. The number of points for this
design was $n=2{k+m\choose m}$, where $m$ was taken at random between $2$ and $7$. 
Output data values consisted of only simulated uniform noise with no specific trend, 
and to each design-output configuration we randomly choose a candidate model $M$ with 
hierarchical structure which was then fitted to data. The size of $M$ determines the 
dimension of $\theta$ and consequently, the complexity of the estimation. 
In total we ran $400$ such scenarios and  
for each scenario, the fit was
performed for both constrained lasso 
and relaxed methodologies as described in Section \ref{sec_estimation}. We generated
a fit using each of the constraints $S,H$ and $W$, using $60$ values of $\lambda$.
For $S$ and $W$ hierarchies we used weights $w_\alpha=10=1/w_\beta$, respectively.
The run time was recorded for each case. In summary, for each of the $400$ simulations, 
	there were $7$ run 
times measured: three for each of $S,H$ and $W$ constraints using constrained lasso; another three for the same
hierarchy cases with relaxed lasso; for comparison we also fitted standard, unconstrained
lasso.

The scatterplot in Figure \ref{fig_runtimes} shows run times for hierarchy $S$ 
against model size. This plot is representative of what happens for the
other cases $W$ and $H$.
The run times for constrained 
lasso are between one and two orders of magnitude bigger than those of relaxed 
constrained lasso. Indeed this is one advantage to be had when using the 
relaxed version of constrained lasso, while
the obvious drawback of it is the potential lack of hierarchy of models, 
as discussed earlier in Section \ref{sec_relaxlasso}.

The increasing pattern of run times is similar in shape for both constrained lasso and relaxed lasso.
This fact is not surprising as both methods consist on quadratic minimization over orthants and what
we are plotting are in essence, runtimes of \texttt{quadprog}.
In both cases, the logarithm of run time appears to depend on the square root of the model size.
Note in the same figure the different pattern and much lower run times of lasso which does not 
depend on constraints hence lasso models are not generally hierarchical. Interestingly enough,
occasionally lasso has similar run times than the constrained versions. 

\subsection{Comparison between methodologies: olive oil data}

We compared the performance of existing methodologies using
the olive oil data set by \cite{FALT1983}. The response 
was the indicator variable for oils coming from the region of Apulia, modelled as a function of 
eight other variables in the dataset. The data set was split randomly in two halves, one was used to
training the model, while the other half of the data was used
to compute the validation error. The smallest validation error in the path was recorded, and 
this procedure was repeated $100$ times for different random splits of the data.

\begin{table}\begin{center}
\begin{tabular}{llll}
Label&Hierarchy&Description and notes\\\hline\hline
S1&Strong& $S$ ($w_\alpha=1$), concides with \cite{BTT2013} for quadratic model\\
H&Strong& $H$&\\
SB&Strong& $S$ ($w_\alpha=|B(\alpha)|$)&\\
SH&Strong & $S$ ($w_\alpha=100$)&\\
WA&Weak &$W$ ($w_\beta=|A(\beta)|$)&\\
WA&Weak &$W$ ($w_\beta=|A(\beta)|$)&\\
W1&Weak &$W$ ($w_\beta=1$)&\\
L&No &Standard lasso of \cite{T1996}\\\hline
ML&No&Lasso for model with main effects only \cite{T1996}\\
R&Strong&RAI by \cite{JSF2019}\\
F&Strong&FAMILY by \cite{HWS2014}\\\hline\hline
\end{tabular}\end{center}
\caption{Labels for scenarios used in the olive oil data.}\label{tab_olive}
\end{table}

\begin{figure}[h]
    \centering
    \includegraphics[scale=0.525]{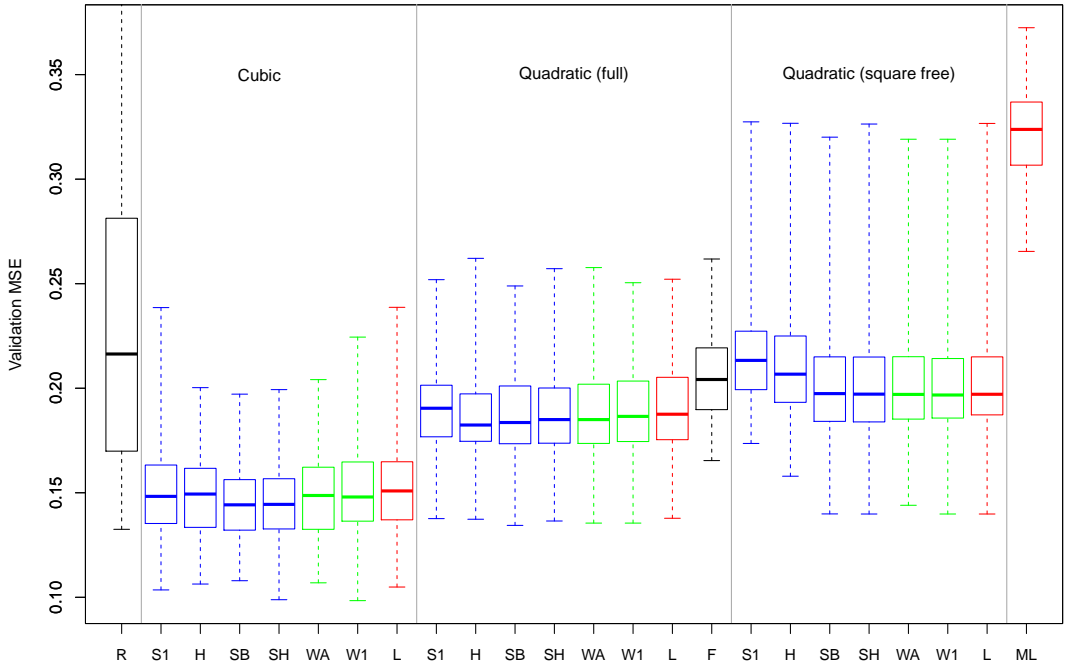}
    \caption{Validation error for different scenarios for the olive oil data.}
    \label{fig_olive}
\end{figure}

The analysis was
carried out for three different candidate models $M$.
The candidate model labelled as ``Quadratic (full)'' consisted of all terms of degree less than or 
equal to two in eight variables totalling $44$ terms. The model labelled as ``Quadratic (square free)'' 
had $36$ terms, obtained by removing the pure terms of degree two from the model ``Quadratic (full)''. 
A more complex model labelled ``Cubic'' with $108$ terms was built adding pure terms of degree three and 
triple interactions to the model ``Quadratic (full)''. To each of these candidate models $M$, seven different 
constraints were tried
to have $21$ scenarios. In addition to these, a simple lasso with main effects only was tried as well
as the proposals RAI and FAMILY, see \cite{JSF2019,HWS2014}.
Boxplots with the results for all these scenarios are shown in Figure \ref{fig_olive}. The boxplots corresponding 
to strong hierarchy $S$ or $H$ are shown in blue, while weak hierarchy $W$ are colored in green, while simple
lasso is shown in red and boxplots in black correspond to other methodologies. Table \ref{tab_olive} contains
a description and labels of the cases that are shown in Figure \ref{fig_olive}.

Overall, adding extra terms of higher degree than two resulted in a model with better prediction
capabilities. Indeed the results for candidate model ``Cubic'' were smaller than the rest of
scenarios. Within this candidate model ``Cubic'', the result labelled `H' in
the figure was the more consistent in the sense of having smaller range of the MSE. The strong 
hierarchy $S$ with
bigger weights, labelled `SH' in the plot, had the smallest errors but also had the second biggest
spread in this group. For the candidate model ``Quadratic (full)'', the result were very
similar amongst methods, with the strong hierarchy labelled as `SB' being slightly better. In this
cathegory note that the method FAMILY with label `F' has slightly worse results.
For candidate model ``Quadratic (square free)'', the results for five of the models are
very similar, with the case `SB' being a little better with smaller range.
Note that for these two cases of quadratic candidate model, the cases `S1' are precisely the
\texttt{hierNet} analyses by \cite{BTT2013}. In the ``Quadratic (full)``, the case `S1' is slightly improved by
`SB', while for ``Quadratic (square free)'' initial model, clearly `S1' is the worst, although
not by much. Finally, we note that apart from a main effects only lasso `ML' which is the second worst
of the methods tried, there is not much difference in each case between lasso and the rest
of methodologies. In other words, adding constraints of the types $S$, $H$ or $W$ 
does not impair severely the predictive capabilities of the model.
The recent methodology RAI searches for models with higher interactions, and while it has the
potential to produce good models, in the examples we tried it had the biggest variability, and
the boxplot shown was trimmed due to its very large right tail.

\section{Discussion}\label{sec_disc}

We perceive there is a lack in the literature for a single, comprehensive approach 
to polynomial hierarchy that is efficient and compares favorably with existing results. 
Our proposal gives direct, intuitive restrictions over the model 
coefficients so that the resulting model satisfies hierarchy. Using this
idea in a Lasso context also provides a simple, efficient search over 
a potentially large set of candidate models. 

The proposed methodology requires knowledge of a candidate model $M$. If this
candidate model is not known, it could be tempting to fit a relatively complex 
hierarchical model and let the constrained lasso procedure determine a suitable model. 
Theoretically this is possible and uses the dual nature of lasso as estimation
and screening procedure. For example, if $M$ is a model of degree two, then our methodology 
coincides with that of \cite{BTT2013}, depending on selected weights. However in general 
we would not advocate to start  
with a complicated model with expensive parameter estimation and we would rather advocate a 
standard two stage procedure. An initial screening
stage would reduce the number of factors and once a reduced set of factors is
available, then a more complex model could be tried. 

A line of future work is concerned with the application and
development of constraints for other statistical models. 
In concrete, a natural development is to adapt our
constrained methodology for the linear predictor $\eta(x)$ of a generalized 
linear model so that $\eta(x)=\sum_{\alpha\in M}\theta_\alpha x^\alpha$. This 
could be particularly advantageous as the linear 
predictor would have more flexibility to describe different, non monotonic
patterns. 

Concerning implementation of our constrained lasso methodology of Section \ref{sec_classo}, work is under progress 
for efficient computation of the lasso path. An initial step
is to take advantage of 
the recently developed package \texttt{quadprogXT}, a development based
upon the techniques of the \texttt{quadprog} library \cite{GI1983,T2013}.

The minimization of the relaxed constrained lasso problem in Section
\ref{sec_relaxlasso} is 
easily implemented using existing the quadratic programming library
\texttt{quadprog} see \cite{T2013}. However, a point that needs
clearer understanding is the matrix in the
the third summand of (\ref{eclassor}). This matrix is at the core 
of the relaxed version of Lasso and is not a full rank matrix.
To overcome this difficulty, a solution 
was to add a multiple of identity to the lower $(2,2)$ block 
of that matrix to have instead
\[\left(\begin{array}{rr}X^TX&-X^TX\\-X^TX&\delta I+X^TX\end{array}\right).\]
This solution has worked well in practice but its stability and accuracy 
needs further study.


\section*{Acknowledgements}
The first author acknowledges partial funding by EPSRC travel grant
EP/K036106/1.

\bibliographystyle{abbrv} 
\bibliography{references}

\newpage

\section*{Appendix}

\subsection*{Proof of Lemma \ref{lemmaSH}}

\begin{proof}
We note that the sum of absolute values 
is greater than or equal than
the any of its absolute components, that is
$\sum_{{\beta'}\in B(\alpha)} |\theta_{\beta'}|\geq|\theta_{\beta}|$ for any ${\beta}\in B(\alpha)$,
and the notation $\beta'$ above is simply to distinguish terms inside the summation.
Hence a consequence of the inequality in Equation (\ref{strongh}) is
that $w_\alpha|\theta_\alpha|\geq |\theta_{\beta}|$ for all
$\beta\in B(\alpha)$, so that when
 $0<w_\alpha\leq 1$ we have $|\theta_\alpha|\geq w_\alpha|\theta_\alpha|\geq |\theta_{\beta}|$.
All the above holds for coefficients of descendants (multiples) $x^\beta$ of $x^\alpha$ 
with exponents ${\beta}\in B(\alpha)$. As
we scan over all constraints of the form (\ref{strongh}) in $S$, we recover all of $H$.
\end{proof}

\subsection*{Proof of Theorem \ref{th_implica}}

\begin{proof}
We first show the implications in the first row involving strong hierarchy 
$S$. To deduce them it is enough to manipulate the 
inequality (\ref{strongh}). 
\begin{description}
\item[S with $w_\alpha<1$ implies S with $w_\alpha=1$.] 
This first implication follows from noting that $w_\alpha<1$ implies that 
$|\theta_\alpha|>w_\alpha|\theta_\alpha|$ so that
the following holds
$|\theta_\alpha|>w_\alpha|\theta_\alpha|\geq \sum_{{\beta}\in B(\alpha)} |\theta_{\beta}|$.
\item[S with $w_\alpha=1$ implies S with $w_\alpha=|B(\alpha)|$.] 
For this implication, we first use Lemma \ref{lemmaSH} to retrieve 
a collection of inequalities of the type $|\theta_\alpha|\geq |\theta_\beta|$ 
for $\beta\in A(\beta)$, that is, constraints from the set $H$. We then add the terms in 
each side of the inequality to retrieve the desired implication. 
\item[S with $w_\alpha=|B(\alpha)|$ implies S with $w_\alpha>|B(\alpha)|$.]
The third and last implication follows from 
noting that if (\ref{strongh}) holds, then
$w_\alpha|\theta_\alpha|>|B(\alpha)||\theta_\alpha|\geq \sum_{{\beta}\in B(\alpha)} |\theta_{\beta}|$
also holds, where $w_\alpha>|B(\alpha)|$.
\end{description}
We now go through the implications in the second row that involve weak hierarchy $W$.
The implications are deduced by manipulating the inequality (\ref{weakh}).

\begin{description}
\item[W with $w_\beta>|A(\beta)|$ implies W with $w_\beta=|A(\beta)|$.] 
This first implication follows from the fact that if 
$w_\beta>|A(\beta)|$, then $w_\beta|\theta_\beta|>|A(\beta)||\theta_\beta|$ so that
$\sum_{{\alpha}\in A(\beta)}|\theta_{\alpha}|\geq w_\beta|\theta_\beta|>|A(\beta)||\theta_\beta|$.
\item[W with $w_\beta=|A(\beta)|$ implies W with $w_\beta=1$.]
This implication is a consecuence of the fact that as $|A(\beta)|\geq 1$ then the
right hand side of inequality obbeys $|A(\beta)||\theta_\beta|\geq |\theta_\beta|$.
\item[W with $w_\beta=1$ implies W with $w_\beta<1$.]
The implication follows from the fact that if (\ref{weakh}) holds for
$w_\beta=1$, then we have 
$\sum_{{\alpha}\in A(\beta)}|\theta_{\alpha}|\geq |\theta_\beta|>w_\beta|\theta_\beta|$
for $w_\beta<1$.
\end{description}

We are only left to show the implications that link sets $S$ and $W$.

\begin{description}
\item[S with $w_\alpha=1$ implies W with $w_\beta=|A(\beta)|$.] 
We first use Lemma \ref{lemmaSH} for all constraints in $S$ with $w_\alpha=1$
to retrieve the full collection of
constraints $H$. Now for every $\beta\in M$ with $A(\beta)\neq\emptyset$,
we add each side of inequalities of the the type $|\theta_\alpha|\geq|\theta_\beta|$
for $\alpha\in A(\beta)$ to retrieve the constraint (\ref{strongh}) with
$w_\beta=|A(\beta)|$.
\item[S with $w_\alpha=|B(\alpha)|$ implies W with $w_\beta=1$.] 
Consider a term $x^\beta$ for which we want to 
determine weak constraints of the type W with $w_\beta=1$. In the development
that 
goes below, refer to this exponent as $\beta'$.
From the set of constraints S with $w_\alpha=|B(\alpha)|$, consider those constraints 
that involve $x^{\alpha}$
for $\alpha\in A(\beta')$, that is
$\left\{|B(\alpha)||\theta_\alpha|\geq \sum_{{\beta}\in B(\alpha)} |\theta_{\beta}|
:\alpha\in A(\beta')\right\}$. As $A(\beta')$ is the set of parent terms to $x^\beta$ then
for each constraint, clearly one of the elements $\beta\in B(\alpha)$ in each sum
is precisely the given $\beta'$
of interest and we obtain the following inequality 
$|B(\alpha)||\theta_\alpha|\geq \sum_{{\beta}\in B(\alpha)} |\theta_{\beta}|\geq |\theta_{\beta'}|$.

In short, we have the collection of inequalities 
\[\left\{|B(\alpha)||\theta_\alpha|\geq |\theta_{\beta'}|:\alpha\in A(\beta')\right\}\] and our task is
to show that these inequalities imply the constraint 
\[\sum_{{\alpha}\in A(\beta')}|\theta_{\alpha}|\geq |\theta_{\beta'}|.\]
The proof is completed by an indirect argument, and we negate the latter constraint
to 
$|\theta_{\beta'}|>\sum_{{\alpha}\in A(\beta')}|\theta_{\alpha}|$
so that $|\theta_\beta|>|\theta_{\alpha}|$ for ${\alpha}\in A(\beta')$.
We immediately
verify that this implies the negation of the former inequalities
$|\theta_{\beta}|>|B(\alpha)||\theta_\alpha|\geq |\theta_\alpha|$ 
for ${\alpha}\in A(\beta')$ which completes the proof.
\end{description}

\end{proof}

\end{document}